\documentclass[10pt,a4paper]{article}

\usepackage{mathptmx}
\usepackage{xspace}
\usepackage[ruled,linesnumbered,vlined]{algorithm2e}
\usepackage{amsmath}
\usepackage[normalem]{ulem}
\usepackage{amssymb}
\usepackage{tikz}

\usepackage{pgfplots}\usetikzlibrary{plotmarks}
\pgfplotsset{
filter discard warning=false 
, legend cell align=left
, minor grid style={loosely dotted, lightgray}
, major grid style={loosely dashed, lightgray}
}

\newtheorem{thm}{Theorem}
\newtheorem{definition}{Definition}
\newtheorem{proposition}{Proposition}
\newtheorem{example}{Example}

\newcommand{\wderives}{\reflectbox{\ \ensuremath{\leadsto}\ }\xspace}

\newcommand{\HT}{{HT}}

\newcommand{\pg}{P_{g}}

\newcommand{\pkg}{\Pi}
\newcommand{\gap}{\mathcal{G}}
\newcommand{\mi}[1]{\ensuremath{\mathit{#1}}}
\newcommand{\mc}{\ensuremath{\mi}{mc}}
\newcommand{\sig}{\Sigma}
\newcommand{\sigk}{\sig^\kappa}

\newcommand{\cS}{{\mathcal F}}

\newcommand{\p}{P}
\newcommand{\pk}{P^\kappa}

\newcommand{\Ik}{I^\kappa}

\newcommand{\tok}[1]{\ensuremath{{#1}^\kappa}}

\newcommand{\toht}[1]{\ensuremath{{#1}^{\HT}}}

\newcommand{\bodyn}[1]{B^-(#1)}
\newcommand{\la}{\leftarrow}

\newcommand{\naf}{\mathop{not}}

\newcommand{\AS}{AS}
\newcommand{\SST}{SST}
\newcommand{\SEQ}{SEQ}

\newcommand{\OAS}{\textit{AS}^{O}}

\newcommand{\NP}{\ensuremath{\mathrm{NP}}}

\newcommand{\SigmaP}[1]{{\Sigma}_{#1}^{P}}
\newcommand{\PiP}[1]{{\Pi}_{#1}^{P}}
\newcommand{\ThetaP}[1]{{\Theta}_{#1}^{P}}

\newcommand{\wasp}{\textsc{wasp}\xspace}

\newcommand{\gringo}{\textsc{gringo}\xspace}

\newcommand{\MM}{\ensuremath{\mathit{MM}}}

\def\<{\langle}
\def\>{\rangle}
\def\int{\mathit{Int}}

\title{On the Computation of Paracoherent Answer Sets\footnote{The original version is published in the Proceedings of the Thirty-First AAAI Conference on Artificial Intelligence (AAAI-17), see \texttt{aaai.org/ocs/index.php/AAAI/AAAI17/paper/download/14358/13877}.}}
\date{}
\author{Giovanni Amendola$^1$, Carmine Dodaro$^1$, Wolfgang Faber$^2$,\\ Nicola Leone$^1$, Francesco Ricca$^1$ \\
         $^1$University of Calabria, Italy\\
         $\{$amendola,dodaro,leone,ricca$\}$@mat.unical.it\\
         $^2$University of Huddersfield, UK\\
         w.faber@hud.ac.uk}

\begin{document}

\maketitle
\label{firstpage}

\begin{abstract}
Answer Set Programming (ASP) is a well-established formalism for nonmonotonic reasoning.
An ASP program can have no answer set due to cyclic default negation. 
In this case, it is not possible to draw any conclusion, even if this is not intended.
Recently, several paracoherent semantics have been proposed that address this issue,
and several potential applications for these semantics have been identified.
However, paracoherent semantics have essentially been inapplicable in practice, due to the lack of efficient algorithms and implementations.
In this paper, this lack is addressed, and several different algorithms to compute semi-stable and semi-equilibrium models are proposed and implemented into an answer set solving framework.
An empirical performance comparison among the new algorithms on benchmarks from ASP competitions is given as well.
\end{abstract}

\section{Introduction}

In the past decades, many advances in Artificial Intelligence research has been done thanks to studies in the field of knowledge representation and reasoning. 
Answer Set Programming (ASP) is a premier formalism for nonmonotonic reasoning
(see, \cite{DBLP:journals/cacm/BrewkaET11,DBLP:series/synthesis/2012Gebser}).
It is a declarative programming paradigm oriented towards difficult, primarily 
NP-hard, search problems, which are encoded into a logic program, 
whose models (answer sets) encode solutions.
ASP is based on the stable model (or answer set) semantics introduced by \cite{gelf-lifs-91}
and is used to solve complex problems in 
Artificial Intelligence~\cite{DBLP:journals/tplp/GagglMRWW15}, 
Bioinformatics~\cite{DBLP:journals/jetai/CampeottoDP15},
Databases~\cite{DBLP:journals/tplp/MannaRT15};
Game Theory~\cite{DBLP:conf/ijcai/AmendolaGLV16}; 
more recently ASP has been applied to solve industrial applications~\cite{DBLP:conf/rr/DodaroLNR15,DBLP:conf/birthday/GrassoLMR11,DBLP:journals/tplp/DodaroGLMRS16}.

However, a logic program could have no answer set due to cyclic default negation. In this case, it is not possible to draw any conclusion, even if this is not intended.
For this reason, theoretical studies have been developed to extend answer set semantics to keep a system responsive in these exceptional cases.
To distinguish this situation from reasoning under classical logical contradiction due to strong negation, called paraconsistent reasoning, it has been referred to it as \textit{paracoherent reasoning} \cite{DBLP:journals/ai/AmendolaEFLM16}.

In order to deal with this, \cite{inou-saka-95} introduced the \textit{semi-stable model semantics} that coincides with answer set semantics whenever a program has some answer set, but admits paracoherent models for each classically consistent program. Recently, \cite{DBLP:journals/ai/AmendolaEFLM16} have improved this kind of semantics avoiding some anomalies with respect to basic modal logic properties, resorting to the equilibrium logic \cite{DBLP:journals/amai/Pearce06}. Thus, this paracoherent semantics is called \textit{semi-equilibrium model semantics}.%
\footnote{The relationship with other semantics is discussed in the Related Work section.}

Different possible applications of these paracoherent semantics have been identified, such as debugging, model building, inconsistency-tolerant query answering, diagnosis, planning and reasoning about actions; and computational complexity aspects have been studied~\cite{DBLP:journals/ai/AmendolaEFLM16}.
However, to the best of our knowledge, there are no efficient algorithms to compute paracoherent answer sets, obstructing the concrete use of these reasoning approaches.
The goal of the paper is to fill this gap, by developing efficient algorithms and their implementation.

In the paper, we consider different algorithms to compute semi-stable and semi-equilibrium models, implementing and integrating them into an answer set building framework. 
Finally, we report the results of an experimental activity conducted on benchmarks from ASP competitions~\cite{DBLP:journals/ai/CalimeriGMR16}, identifying the more efficient algorithm.


\section{Preliminaries}
We start with recalling answer set semantics, 
and then present the paracoherent semantics of 
semi-stable and semi-equilibrium models.

\subsection{Answer Set Programming}
We concentrate on programs over a propositional signature $\sig$.
A \textit{disjunctive rule} $r$ is of the form 
\begin{equation}
 a_{1}\vee\cdots\vee a_{l} \leftarrow b_{1},...,b_{m},not \
 b_{m+1},...,not \ b_{n},
\label{eq:rule}
\end{equation}
\noindent where all $a_i$ and $b_j$ are atoms (from $\sig)$ and $l\geq 0$,
$n\geq m\geq 0$ and $l+n>0$; $\naf$ represents
\textit{negation-as-failure}. The set $H(r)=\lbrace
a_{1},...,a_{l} \rbrace$ is the \textit{head} of $r$, while $B^{+}(r)=\lbrace b_{1},...,b_{m} \rbrace$ and $B^{-}(r)=\lbrace
b_{m+1},\ldots,b_{n} \rbrace$ are
the \textit{positive body} and the \textit{negative body} of $r$,
respectively; the \textit{body} of $r$
is $ B(r)=B^{+}(r)\cup B^{-}(r)$. We denote by $At(r)=H(r)\cup B(r)$ the
set of all atoms occurring in $r$. 
%
A rule $r$ is a \textit{fact}, if $B(r)=\emptyset$ (we then omit
$\leftarrow$); a \textit{constraint}, if $H(r)=\emptyset$;
\textit{normal}, if $| H(r)| \leq 1$ and \textit{positive}, if
$B^{-}(r)=\emptyset$. 
A \textit{(disjunctive logic) program} $P$ is a
finite set of disjunctive rules. $P$ is called \textit{normal}
[resp.\ \textit{positive}] if each $r\in P$ is normal
[resp.\ positive]. We let $At(P)=\bigcup_{r\in P} At(r)$.

Any set $I\subseteq \sig$ is an \textit{interpretation}; 
it is a \textit{model} of a program $P$ (denoted
$I\models P$) iff for each rule $r\in P$, $I\cap H(r)\neq \emptyset$ if
$B^{+}(r)\subseteq I$ and $B^{-}(r)\cap I=\emptyset$ (denoted $I \models
r$).  A model $M$ of $P$ is \textit{minimal}, iff no model $M'\subset M$ of $P$
exists.
We denote by $\MM(P)$ the set of all minimal
models of $P$ and by $AS(P)$ the set of all {\em answer sets (or stable
  models)}\/ of $P$, i.e., the set of all interpretations  $I$ such that
$I\in \MM(P^I)$, where $P^I$ is the
well-known \textit{Gelfond-Lifschitz reduct} \cite{gelf-lifs-91} of
$P$ with respect to $I$,
 i.e., the set of rules
$ a_{1}\vee ...\vee a_{l} \leftarrow b_{1},...,b_{m}$,
obtained from rules $r\in P$ of form (\ref{eq:rule}), such that
$B^{-}(r)\cap I=\emptyset$.
A program is said to be coherent if $AS(P) \neq \emptyset$, incoherent otherwise.

Now, we recall a useful extension of the answer set semantics by the notion of {\em weak constraint}~\cite{DBLP:journals/tkde/BuccafurriLR00}.
A weak constraint $\omega$ is of the form:
\begin{center}
	$\wderives b_1,\ldots, \ b_m, \ \naf b_{m+1},\ldots, \ \naf b_n.$
\end{center}
%
Given a program $P$ and a set of weak constraints $W$, the semantics of $P\cup W$ extends from the basic case defined above. A constraint $\omega \in W$ is violated by an interpretation $I$ if all positive atoms in $\omega$ are true, and all  negated atoms are false with respect to $I$.
An {\em optimum answer set} for $P\cup W$ is an answer set of $P$ that minimizes
the number of the violated weak constraints.
We denote by $\OAS(P\cup W)$ the set of all optimum answer sets of $P\cup W$.

\subsection{Paracoherent ASP}

Here, we introduce two paracoherent semantics that allow for keeping a system responsive when a logic program has no answer set due to cyclic default negation. These semantics satisfy three desiderata properties identified by~\cite{DBLP:journals/ai/AmendolaEFLM16}.
\paragraph{Semi-Stable Models.}
Inoue and Sakama (\cite{inou-saka-95}) introduced  \emph{semi-stable model semantics}.
We consider an extended signature
$\sigk = \sig\cup\{Ka\mid a\in \sig\}$.
Intuitively, $Ka$ can be read as $a$ is believed to hold. 
Semantically, we resort to subsets of $\sigk$ as interpretations $\Ik$ and the truth values 
false $\bot$, believed true $\mathbf{bt}$, and true $\mathbf{t}$. 
The truth value assigned by $\Ik$ to a propositional variable $a$ is defined by 
$$\Ik(a) =\left\{\begin{array}{ll}
 \mathbf{t} & \mbox{if}\  a \in \Ik, \\[0.5ex]
 \mathbf{bt} & \mbox{if}\ Ka \in \Ik\ \mbox{and}\ a \not\in \Ik,\\[0.5ex]
 \bot & \mbox{otherwise}.
\end{array}\right.$$
The semi-stable models of a program $\p$ are obtained from its \emph{epistemic $\kappa$-transformation} $\pk$.
\begin{definition}
[Epistemic $\kappa$-transformation $\pk$]\label{def:epi-trans}
Let $\p$ be a program. Then its epistemic $\kappa$-transformation is defined as 
the program $\pk$ obtained from $\p$ by replacing each rule $r$  
of the form~(\ref{eq:rule}) in $\p$, such that $\bodyn{r}\neq\emptyset$, with:
\begin{align}
\lambda_{r,1} \vee \ldots \vee \lambda_{r,l} \vee Kb_{m+1} \vee \ldots \vee Kb_n & \la  b_1,\ldots, b_m, \\ 
a_i & \la \lambda_{r,i}, \\ 
    & \la \lambda_{r,i}, b_j, \\
\lambda_{r,i} & \la  a_i, \lambda_{r,k},
\end{align}
for $1\leq i,k\leq l$ and $m+1\leq j\leq n$, where the $\lambda_{r,i}$, $\lambda_{r,k}$ are fresh atoms.
\end{definition}
Note that for any program $\p$, its epistemic $\kappa$-transformation $\pk$ is positive.  
%
For every interpretation $\Ik$ over $\sig'\supseteq\sigk$, let $\gap(\Ik)=\{ Ka\in\Ik\ \mid a\not\in\Ik\}$ denote the atoms believed true but not assigned true, also referred to as the gap of $\Ik$.
Given a set $\cS$ of interpretations over $\sig'$, an interpretation $\Ik\in \cS$ is \emph{maximal canonical in $\cS$}, if no $\tok{J}\in \cS$ exists such that 
$\gap(\Ik)\supset\gap(\tok{J})$.
By $\mc(\cS)$ we denote the set of maximal canonical interpretations in $\cS$.
Semi-stable models are then defined as \emph{maximal canonical} interpretations among the answer sets of $\pk$. 
Then we can equivalently paraphrase the definition of semi-stable models in~\cite{inou-saka-95} as follows.

\begin{definition}[Semi-stable models]\label{def:sst}
Let $\p$ be a program over $\sig$. 
An interpretation $\Ik$ over $\sigk$ is a \emph{semi-stable model} of $\p$, 
if $\Ik = S\cap \sigk$ for some maximal canonical answer set $S$ of $\pk$.
The set of all semi-stable models of $\p$ is denoted by $\SST(\p)$,
i.e., $\SST(\p)= \{  S \cap \sigk \mid S\in\mc(\AS(\pk))\}$.
\end{definition}

\begin{example}\label{ex:sst}
Consider the program $\p=\{b\la\naf a; \ c\la\naf b; \ a\la c; \ d\la\naf d \}$.  
Its epistemic $\kappa$-transformation is 
$\pk = \{  
\lambda_1 \lor Ka;\ b \la \lambda_1;\  \la a, \lambda_1;\ \lambda_1 \la b,\lambda_1; \
\lambda_2 \lor Kb;\ c \la \lambda_2;\  \la b, \lambda_2;\ \lambda_2 \la c,\lambda_2;
\ a\la c; \
\lambda_3 \lor Kd;\ d \la \lambda_3;\  \la d, \lambda_3;\ \lambda_3 \la d,\lambda_3;
  \}$, 
which has the answer sets 
$M_1 = \{ Ka, Kb, Kd \}$, 
$M_2= \{\lambda_1, b, Kb, Kd\}$, and
$M_3= \{Ka, \lambda_2, a, c, Kd\}$; 
as $\gap(M_1) = \{Ka, Kb, Kd\}$, $\gap(M_2) = \{ Kd\}$,  
and $\gap(M_3) =\{ Kd\}$. 
Therefore, among them $M_2$ and $M_3$ are maximal canonicals, 
and hence $M_2\cap\sigk = \{ b, Kb, Kd \}$ 
and $M_3\cap\sigk = \{a,c,Ka, Kd \}$
are semi-stable models of $\p$,
that also correspond to answer sets of $\p$. 
\end{example}

\paragraph{Semi-Equilibrium Models.}
Semi-equilibrium models were introduced by~\cite{DBLP:journals/ai/AmendolaEFLM16}
to avoid some anomalies in semi-stable model semantics.
Like semi-stable models, semi-equilibrium models may be computed as maximal canonical 
answer sets, of an extension of the epistemic $\kappa$-transformation. 

\begin{definition}[Epistemic $\HT$-transformation $\toht{\p}$] 
\label{def:ht-trans}
Let $\p$ be a program over $\sig$. Then its epistemic $\HT$-transformation 
$\toht{\p}$ is defined as the union of $\pk$ with the set of rules:
\begin{align}
Ka & \la a,\\
Ka_1 \vee \ldots \vee Ka_l \vee Kb_{m+1} \vee \ldots \vee Kb_n 
&\la  Kb_1,\ldots, Kb_m, 
\end{align}
\noindent for $a \in \sig$, respectively for every rule $r\in\p$ of the form~(\ref{eq:rule}).
\end{definition}

\begin{definition}[Semi-equilibrium models]\label{def:seq}
Let $\p$ be a program over $\sig$, and let $\Ik$ be an interpretation 
over $\sigk$. 
Then, $\Ik\in\SEQ(\p)$
if, and only if, $\Ik\in\{ M\cap\sigk \mid M\in\mc(\AS(\toht{\p}))\}$,
where $\SEQ(\p)$ is the set of semi-equilibrium models of $\p$.
\end{definition}

\begin{example}\label{ex:seq}
Consider the program $\p$ of Example~\ref{ex:sst}.  
Its epistemic $\HT$-transformation $\toht{\p}$ is 
$\pk \cup \{  
Ka \la a; \ Kb \la b; \ Kc\la c; \ Kd\la d; \
Kb \vee Ka; \ 
Kc \vee Kb; \
Ka \la  Kc; \
Kd \la  Kd  \}$, 
which has the answer sets 
$\{ Ka, Kb, Kd \}$, 
$\{\lambda_1, b, Kb, Kd\}$, and
$\{Ka, \lambda_2, a, c, Kc, Kd\}$.
Therefore, the semi-equilibrium models of $\p$ are
$\{ b, Kb, Kd \}$ 
and $\{a,c,Ka,Kc, Kd \}$. 
\end{example}

In the following, we refer to semi-stable models or semi-equilibrium models as \textit{paracoherent answer sets}.

\paragraph{Complexity Considerations.}
The complexity of various reasoning tasks with paracoherent answer
sets has been analyzed in \cite{DBLP:journals/ai/AmendolaEFLM16}:
while determining the existence of paracoherent answer sets is
$\NP$-complete (it is sufficient to test for existence of classical
models), paracoherent answer set checking is $\PiP{2}$-complete,
leading to $\SigmaP{3}$-completeness for brave, and
$\PiP{3}$-completeness for cautious reasoning. In this paper, we
consider the computation of one paracoherent answer set, which is 
a functional problem. From previous work it is clear that this task is
in F$\SigmaP{3}$, and actually in F$\ThetaP{3}$ (functional polynomial
time with a logarithmic number of calls to a $\SigmaP{2}$-complete
oracle), because for computing one paracoherent answer set it is
sufficient to solve a cardinality-optimization problem.



\begin{algorithm}[t!]
	$M := nextAnswerSet(\pkg,\perp)$; \qquad $M^w := M$\; $M^w :=
	nextAnswerSet(\pkg,M^w)$\; \label{Alg:filter:ln:next} \lIf{$M^w=\perp$}{\Return{$M$}}\; \lIf{$gap(M^w) \subset
	gap(M)$}{$M :=
	M^w$}\; \textbf{goto}~\ref{Alg:filter:ln:next}; \caption{Filtering} \label{Alg:filter}
\end{algorithm}

\begin{algorithm}[t!]
	$M = \perp$\;
	$M = nextAnswerSet(\pkg,M)$\; \label{Alg:GC:ln:next}
	$M^{w} = nextAnswerSet(\pkg\cup \pkg_{M},\perp)$\;
	\lIf{$M^w = \perp$}{\Return{$M$};}
	\lElse{\textbf{goto}~\ref{Alg:GC:ln:next};}
	\caption{Guess\&Check}
	\label{Alg:GC}
\end{algorithm}

\section{Computation of a Paracoherent Answer Set}
In this section we propose different algorithms to compute one paracoherent answer set.
The algorithms take as input a program $\pkg=P^\chi \cup \pg$, where $P^\chi$ is a generic epistemic transformation of the ASP program $P$ and  $\pg$ is the following set of rules capturing the notion of gap:
\begin{align}
gap(Ka) & \la Ka, \ \naf a; & \forall a\in At(P)
\label{eq:rule:Gap}
\end{align}
\begin{proposition}
Let $gap(I) = \{gap(Ka) \mid gap(Ka) \in I\}$, for a set $I$ of atoms.
An answer set $M$ of $\pkg$ is a paracoherent answer set if, and only if, there exists no answer set $M_1$ of $\pkg$ such that $gap(M_1) \subset gap(M)$.
\end{proposition}

\begin{example}\label{ex:pkg}
	Consider again the program $\pk$ of Example~\ref{ex:sst}, then $\pkg$ is the union of $\pk$ with the following set of rules:
	$$
	\begin{array}{cc}
		gap(Ka) \la Ka,\  \naf a; & gap(Kb) \la Kb,\  \naf b; \\
		gap(Kc) \la Kc,\  \naf c;  & gap(Kd) \la Kd,\  \naf d; \\
	\end{array}
	$$
	which admits the answer sets $M_1' = M_1 \cup\{gap(Ka), gap(Kb), gap(Kd)\}$,  $M_2' = M_2 \cup \{gap(Kd)\}$, and $M_3' = M_3 \cup \{gap(Kd)\}$.
	Then, $gap(M_1')=\{gap(Ka), gap(Kb), gap(Kd)\}$, $gap(M_2')=gap(M_3')=\{gap(Kd)\}$, thus $M_2'$ and $M_3'$ are paracoherent answer sets.
\end{example}

The output of the algorithms is one semi-stable model of $P$ (if $\chi=\kappa$) or one semi-equilibrium model of $P$ (if $\chi=\HT$).
In the following, without loss of generality we assume that $\pkg$ admits at least one paracoherent answer set.
In fact, by properties of semi-stable and semi-equilibrium models, this kind of programs admit always a paracoherent answer set \cite{DBLP:journals/ai/AmendolaEFLM16}.

Moreover, in order to ease the description of the algorithms presented in this section, we introduce the enumeration function $nextAnswerSet$, that takes as input the program $\pkg$ and an answer set $M$ of $\pkg$, and returns as output the next one according to some internal criteria or $\perp$ if no other answer set exists. We abuse of the notation using $M=\perp$ to indicate that the function computes the first answer set.

\begin{algorithm}[t!]
	$M := nextAnswerSet(\pkg,\perp)$\;
	$\pkg := \pkg\cup \pkg_{M}$\; \label{Alg:minimize:ln:prog}
	$M^{w} := nextAnswerSet(\pkg,\perp)$\; \label{Alg:minimize:ln:search}
	\lIf{$M^w = \perp$}{\Return{$M$};}
	\lElse{$M := M^{w}$; \textbf{goto}~\ref{Alg:minimize:ln:prog};}
	\caption{Minimize}
	\label{Alg:minimize}
\end{algorithm}

\begin{algorithm}[t!]
	$M := nextAnswerSet(\pkg,\perp)$; \qquad $C:= gap(M)$\;
	\lIf{$C=\emptyset$\label{Alg:split:ln:check}}{\Return $M$}\;
	$\pkg := \pkg\cup \pkg_{M}$; \qquad $a := OneOf(C)$\;
	$M^{w} := nextAnswerSet(\pkg \cup \{\la a\},\perp)$\; \label{Alg:split:ln:next}
	\lIf{$M^w = \perp$}{\{$\pkg := \pkg \cup \{\la \naf a\}$; $C := C \setminus \{a\};$\}}
	\lElse{\{$M := M^{w}$; $C:= gap(M^w);\}$}
	\textbf{goto}~\ref{Alg:split:ln:check};
	\caption{Split}
	\label{Alg:split}
\end{algorithm}

\paragraph{Filtering.}
An immediate algorithm for finding a paracoherent answer set is \emph{Filtering}, Algorithm~\ref{Alg:filter}.
The underlying idea is to enumerate all answer sets of $\pkg$ and to store the one that is subset-minimal with respect to gap atoms.
The algorithm first finds an answer set $M$ of $\pkg$.
Then, another answer set $M^w$ is searched (line~\ref{Alg:filter:ln:next}).
If $gap(M^w)$ is a subset of $gap(M)$ then $M$ is replaced with $M^w$.
Subsequently, the algorithm continues the search until all answer sets have been enumerated.
Intuitively, at each step of the computation $M$ is a subset-minimal answer set with respect to the answer sets enumerated so far.
Thus, when all answer sets have been enumerated then $M$ is a paracoherent answer set.
\begin{example}
Consider again program $\pkg$ of Example~\ref{ex:pkg}.
The first call to $nextAnswerSet$ returns $M_1'$ that is stored in $M$.
The second call to $nextAnswerSet$ returns $M_2'$, $gap(M_2')$ is a subset of $gap(M_1')$ therefore $M$ is replaced by $M_2'$.
The third call of $nextAnswerSet$ returns $M_3'$ and $M$ is not modified since $gap(M_3')'$ is not a subset of $gap(M)$.
No other answer sets can be enumerated, thus the algorithm terminates returning $M$.
\end{example}

The main drawback of Algorithm~\ref{Alg:filter} is that it always computes all answer sets of $\Pi$, a potentially exponential number in the size of the atoms of the original program.

In the following we present different algorithms for addressing this inefficiency.

\begin{table*}[t!]
	\centering
\scriptsize{
	\begin{tabular}{|lr||rrrrr||rrrrr|}
		\cline{1-12}
		\rule{0pt}{2.3ex}&&	\multicolumn{5}{c||}{Semi-stable semantics} & \multicolumn{5}{c|}{Semi-equilibrium semantics}\\
		\cline{3-12}
		\rule{0pt}{2.3ex}\textbf{Problems}	&	\textbf{\#} &	\textbf{Filt}	&	\textbf{G\&C}	&	\textbf{Minim}	&	\textbf{Split}	&	\textbf{Weak}	&	\textbf{Filt}	&	\textbf{G\&C}	&	\textbf{Minim}	&	\textbf{Split}	&	\textbf{Weak}\\
		\cline{1-12}
		\rule{0pt}{2.3ex}KnightTour				&	26	   &	0	&	0	&	0	&	0	&	0	&		0	&	0	&	0	&	0	&	0\\
		MinDiagn				&	64	   &	0	&	37	&	47	&	47	&	0	&		0	&	0	&	8	&	13	&	0\\
		QualSpatReas		&	61	  &	0	&	26	&	26	&	26	&	10	&		0	&	0	&	0	&	0	&	0\\
		StableMarriage		&	1		&	0	&	0	&	0	&	0	&	0	&		0	&	0	&	0	&	0	&	0\\
		VisitAll					&	 5		  &	0	&	5	&	5	&	5	&	5	&		0	&	5	&	5	&	5	&	5\\
		\cline{1-12}
		\rule{0pt}{2.3ex}\textbf{Total}						&	\textbf{157} 	&	\textbf{0}	&	\textbf{68}	&	\textbf{78}	&	\textbf{78}	&	\textbf{15}	&		\textbf{0}	&	\textbf{5}	&	\textbf{13}	&	\textbf{18}	& \textbf{5}\\
		\cline{1-12}
	\end{tabular}}
	\scriptsize{
	\caption{Number of instances solved within the allotted time.}		
	}\label{tab:results}
\end{table*}


\begin{table*}[t!]
	\centering
	\scriptsize{
	\begin{tabular}{|lr||rr||rr||rr|}
		\cline{1-8}
		\rule{0pt}{2.3ex}&&	\multicolumn{2}{c||}{$P$} &	\multicolumn{2}{c||}{$\pk \cup \pg$} & \multicolumn{2}{c|}{$\toht{\p} \cup \pg$}\\
		\cline{3-8}
		\rule{0pt}{2.3ex}\textbf{Problems}	&	\textbf{\#} & \textbf{avg \#atoms} &\textbf{avg \#rules} & \textbf{avg \#atoms} & \textbf{avg \#rules} & \textbf{avg \#atoms} & \textbf{avg \#rules}\\
		\cline{1-8}
		\rule{0pt}{2.3ex}KnightTour				&	26	   & 101\,480 & 474\,852 & 358\,035	& 655\,709	& 358\,035	& 1\,238\,936 \\
		MinDiagn				&	64	   & 238\,677 & 654\,412 & 1\,214\,042 &	1\,579\,409	& 1\,214\,042	& 2\,549\,729\\
		QualSpatReas		&	61	  & 11\,774 & 735\,317 & 68\,793	& 789\,642	& 68\,793 &	1\,542\,607 \\
		StableMarriage		&	1		& 649\,326 & 110\,536\,185 & -	& - & 	-	& - \\
		VisitAll					&	 5		  & 3\,820 & 64\,234 & 13\,926 &	72\,776	& 13\,926 &	140\,881 \\
		\cline{1-8}
	\end{tabular}
	\caption{Impact of epistemic transformations $\pk$ and $\toht{\p}$.}		
	\label{tab:rewriting}}	
\end{table*}

\paragraph{Guess\&Check.} 
This algorithm, Algorithm~\ref{Alg:GC}, improves Algorithm~\ref{Alg:filter} by reducing the number of computed answer sets.
In order to ease the description of the remaining algorithms we introduce the following.
\begin{definition}
	Given a program $\pkg$ defined as above. 
	Let $M$ be a model of $\pkg$, then $\pkg_M$ is the following set of constraints:
	\begin{align}
	& \la gap(M); \label{cons:gapM}\\
	& \la gap(Ka); & \forall gap(Ka) \in At(\pkg) \setminus M. \label{eq:gapconstraints}
	\end{align}
Note that (\ref{cons:gapM}) contains all atoms in $gap(M)$.
\end{definition}

\begin{thm}\label{th:Disj}
Let $P$ be a logic program, let $\pkg$ be defined as above, and let $M\in\AS(\pkg)$.
Then, $\AS(\pkg\cup \pkg_M)\neq\emptyset$ if, and only if, $M$ is not a paracoherent answer set of $P$.
\end{thm}

\begin{example}
	Consider again program $\pkg$ of Example~\ref{ex:pkg}.
	$\pkg_{M_1'}$ is composed by the following set of constraints: 
	$$
	\begin{array}{lr}
	\la gap(Ka), gap(Kb), gap(Kd); &  \la gap (Kc);
	\end{array}
	$$
	whereas $\pkg_{M_2'}$ is composed by the following set of constraints:
	$$
	\begin{array}{cc}
	\la gap(Kd); & \la gap(Ka);	\\ \la gap(Kb); & \la gap(Kc).\\
	\end{array}
	$$
	Note that $\AS(\pkg \cup \pkg_{M_1'}) = \{M_2',M_3'\}$ and $\AS(\pkg \cup \pkg_{M_2'}) = \emptyset$.
\end{example}

The Guess\&Check algorithm finds an answer set $M$ of $\pkg$.
Subsequently, an answer set of the program $\pkg \cup \pkg_{M}$ is sought.
If such an answer set does not exist then $M$ is a paracoherent answer set and the algorithm terminates returning $M$.
Otherwise, the algorithm iterates the computation until a paracoherent answer set is found.
\begin{example}
	Consider again program $\pkg$ of Example~\ref{ex:pkg}.
	The first answer set computed by $nextAnswerSet$ is $M_1'$. The subsequent check is performed on the program $\pkg \cup  \pkg_{M_1'}$, that is coherent. Thus, $M_1'$ is not a paracoherent answer set.
	Then, $nextAnswerSet$ is called again and it returns $M_2'$. At this point, $\pkg \cup \pkg_{M_2'}$ is incoherent, therefore the algorithm terminates returning $M_2'$.
\end{example}

Algorithm~\ref{Alg:GC} terminates as soon as a paracoherent answer set of $P$ is found. However, in the worst case, it still needs to enumerate all answer sets.




\paragraph{Minimize.}
The next algorithm is called Minimize and it is reported as Algorithm~\ref{Alg:minimize}.
The idea is to compute an answer set $M$ of $\pkg$ and then to search for another answer set $M^w$ such that $gap(M^w) \subset gap(M)$.
This property is enforced by the constraints of $\pkg_{M}$ that are added to the program $\pkg$ (line~\ref{Alg:minimize:ln:prog}).
If $\pkg$ admits an answer set, say $M^w$, then $M$ is replaced by $M^w$ and the algorithm iterates minimizing $M$.
Otherwise, if $\pkg$ admits no answer set, $M$ is a paracoherent answer set and the algorithm terminates returning $M$.

\begin{example}
	Consider again program $\pkg$ of Example~\ref{ex:pkg}.
	The first answer set computed by $nextAnswerSet$ is $M_1'$. Thus, the constraints of $\pkg_{M_1'}$ are added to $\pkg$.
	The subsequent check on $\pkg$ returns an answer set, say $M_2'$, and then $\pkg$ is modified by adding the constraints of $\pkg_{M_2'}$.
	At this point, $\pkg$ is incoherent, therefore the algorithm terminates returning $M_2'$.
\end{example}

Algorithm~\ref{Alg:minimize} computes at most $|At(P)|$ answer sets.

\paragraph{Split.}
Another algorithm for computing a paracoherent answer set is called Split, Algorithm~\ref{Alg:split}.
The algorithm first computes an answer set $M$ of $\pkg$ and creates a set $C$ of gap atoms that are included in $M$.
Then, the program $\pkg$ is modified by adding the constraints of $\pkg_{M}$.
Moreover, one of the atoms in $C$ is selected by the procedure $OneOf$, say $a$.
Subsequently, an answer set of $\pkg \cup \{\la a\}$ is searched.
If such an answer set does not exist then $a$ must be included in the paracoherent answer set and thus $\pkg$ is modified by adding the constraint $\la \naf a$ and $a$ is removed from the set $C$.
Otherwise, if $\pkg \cup \{\la a\}$ admits an answer set, say $M^w$, then $M$ is replaced by $M^w$ and the set $C$ is replaced by the gap atoms  that are true in $M^w$.
The algorithm then iterates until the set $C$ is empty, returning $M$ that corresponds to the paracoherent answer set.

\begin{example}
	Consider again program $\pkg$ of Example~\ref{ex:pkg}.
	The first answer set computed by $nextAnswerSet$ is $M_1'$. Thus, $C$ is set to $\{gap(Ka), gap(Kb), gap(Kc)\}$ and the constraints of $\pkg_{M_1'}$ are added to $\pkg$.
	Then, function $OneOf$ selects one of the atoms in $C$, say $gap(Ka)$.
	The subsequent check on $\pkg \cup \{\la gap(Ka)\}$ returns an answer set, say $M_2'$.
	Therefore, $C$ is set to $gap(Kd)$ and $\pkg$ is modified by adding the constraints of $\pkg_{M_2'}$.
	Then, the function $OneOf$ selects $gap(Kd)$ and the subsequent check on $\pkg \cup \{\la gap(Kd)\}$ returns $\perp$.
	Subsequently, $\pkg$ is modified by adding the constraint $\la \naf gap(Kd)$ and $C$ is updated by removing $gap(Kd)$.
	At this point, $C$ is empty, therefore the algorithm terminates returning the latest computed answer set, i.e. $M_2'$.
\end{example}

Note that Algorithm~\ref{Alg:split} requires to compute at most $|At(P)|$ answer sets.

\paragraph{Weak constraints.}
All the algorithms presented above require the modification of an ASP solver to be implemented.
An alternative approach is based on the observation that the gap minimality can be obtained adding to $\pkg$ the following set of weak constraints, say $W$:
\begin{align}
& \wderives gap(Ka); & \forall a\in At(P).
\end{align}
The answer set of the extended program is then an answer set of $\pkg$ such that a minimal number of weak constraints in $W$ is violated. This means that this answer set that is cardinality minimal with respect to the gap atoms.
Therefore, it is also subset minimal with respect to the gap atoms, and so, it is a paracoherent answer set of $P$.

\begin{thm}
Let $P$ be a program, let $\pkg$ and $W$ be defined as above.
If $M\in \OAS(\pkg\cup W)$, then $M\setminus gap(M)$ is a paracoherent answer set of $P$. 
\end{thm}

Note that, the reverse statement does not hold in general.
For example, consider the program 
$P=\{b\la \naf a; \ c\leftarrow a; \ d\leftarrow b, \naf d\}$. 
Its semi-equilibrium models are $\{b,Kd\}$ and $\{Ka,Kc\}$.
However, $\{Ka,Kc,gap(Ka),gap(Kc)\}$ is not an optimum answer set of $\pkg\cup W$. 


%



\section{Implementation and Experiments}

We implemented the algorithms presented in this paper, and we report on an experiment comparing their performance.

\paragraph{Implementation.}
The computation of a paracoherent answer set is obtained in two steps.
First a Java rewriter computes the epistemic transformations $\pk$ and $\toht{\p}$ of a propositional ASP program. Then the output of the rewriter is fed in input to a variant of the state-of-the-art ASP solver \wasp~\cite{DBLP:conf/lpnmr/AlvianoDLR15}.
\wasp is an open-source ASP solver, winner of the latest ASP competition~\cite{DBLP:conf/lpnmr/GebserMR15}, that we modified by implementing the algorithms presented in the previous section (the source can be downloaded at \textit{https://github.com/alviano/wasp}).%
%

\paragraph{Benchmarks settings.}
Experiments were run on a Debian Linux system with 2.30GHz Intel Xeon E5-4610 v2 CPUs and 128GB of RAM. Execution time and memory were limited to 1200 seconds and 3 GB, respectively.
We use benchmark instances from the latest ASP competition~\cite{DBLP:conf/lpnmr/GebserMR15} collection. We consider all the incoherent instances that do not feature in the encoding neither \textit{aggregates}, nor \textit{choice rules}, nor \textit{weak constraints}, since such features are not currently supported by the paracoherent semantics~\cite{DBLP:journals/ai/AmendolaEFLM16}. 
This resulted in instances from the following domains: Knight Tour, Minimal Diagnosis, Qualitative Spatial Reasoning, Stable Marriage and Visit All.
Instances were grounded with \gringo (from \textit{http://potassco.sourceforge.net/}).
Grounding times, the same for all compared methods, are not reported.

\paragraph{Results of the experiments.}
A summary of the result is reported in Table~\ref{tab:results}, where the number of solved instances for each considered semantics  is reported.
In the table, Filt is \wasp running Algorithm~\ref{Alg:filter}, G\&C is \wasp running Algorithm~\ref{Alg:GC}, Minim is \wasp running Algorithm~\ref{Alg:minimize}, Split is \wasp running Algorithm~\ref{Alg:split}, and Weak is \wasp running the algorithm based on weak constraints.

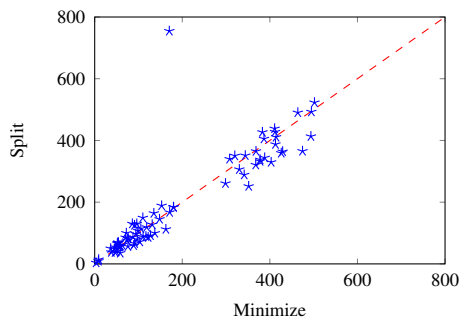
\begin{figure}[t!]
\centering
\begin{tikzpicture}
\pgfkeys{%
/pgf/number format/set thousands separator = {}}
\begin{axis}[
scale only axis
, font=\scriptsize
, x label style = {at={(axis description cs:0.5,0.04)}}
, y label style = {at={(axis description cs:0.05,0.5)}}
, xlabel={Minimize}
, ylabel={Split}
, width=0.38\textwidth
, height=0.27\textwidth
, xmin=0, xmax=800
, ymin=0, ymax=800
, xtick={0,200,400,600,800}
, ytick={0,200,400,600,800}
, major tick length=2pt
, title={}
]

\addplot [mark size=2pt, only marks, color=blue, mark=star, dashed] [unbounded coords=jump] table[x index=1, y index=2] {\results}; 	
\addplot [color=red, dashed] [unbounded coords=jump] table[x index=0, y index=0] {\results}; 
\end{axis}
\end{tikzpicture}
\caption{Instance by instance comparison between Minimize and Split on semi-stable semantics.}\label{fig:scatter}
\end{figure}

As a general comment, the algorithm based on the enumeration of answer sets is highly inefficient solving no instances at all.
The Guess\&Check algorithm outperforms the Filtering algorithm demonstrating that in many cases the enumeration of all answer sets is not needed. 
The best performing algorithms are Minimize and Split solving both the same number of instances.
The performance of the two algorithms are similar also considering the running times.
In fact, this is evident by looking at the instance-wise comparison reported in the scatter plot of Figure~\ref{fig:scatter}.
A point \textit{(x,y)} in the scatter plot is reported for each instance, where \textit{x} is the solving time of the algorithm Minimize whereas \textit{y} is the solving time of the algorithm Split.
Concerning the algorithm based on weak constraints, it can be observed that its performance is better than the one of algorithm Filtering.
However, it does not reach the efficiency of the algorithms Guess\&Check, Minimize and Split.

Concerning the semi-equilibrium semantics, it can be observed that the performance of all algorithms deteriorates. 
This can be explained by looking at the number of rules introduced by the epistemic \HT-transformation, reported in Table~\ref{tab:rewriting}.
In fact, the epistemic \HT-transformation introduces approximately twice the number of rules introduced by the epistemic $\kappa$-transformation.
Moreover, we observe that also in this case the best performing algorithms are Minimize and Split. The latter is slightly more efficient 
than the former.

Focusing on the performance of the algorithms on the different benchmarks, it can be observed that none of the algorithms was effective on the problems KnightTour and StableMarriage.
Concerning KnightTour, we observed that \wasp is not able to find any answer set of the epistemic transformations for 12 out of 26 instances. 
Basically, no algorithm can be effective on such 12 instances, and
the remaining ones are hard due to the subsequent checks.
Concerning StableMarriage, we observed that java rewriter could not produce the epistemic transformation within the allotted time, because the unique instance of this domain features more than 100 millions rules.
The presented algorithms (but Filtering) are able to solve all the considered instances of VisitAll problem, where the epistemic transformations results in a very limited number of atoms and rules (see Table~\ref{tab:rewriting}).

\pgfplotstableread[row sep=\\, col sep=&]{
	1 & 44.9 & 48.97\\
	2 & 47.34 & 53.87\\
	3 & 38.03 & 37.59\\
	4 & 49.24 & 39.18\\
	5 & 36.5 & 49.81\\
	6 & 58.65 & 58.47\\
	7 & 51.7 & 66.03\\
	8 & 56.87 & 58.75\\
	9 & 62.14 & 61.62\\
	10 & 54.93 & 55.04\\
	11 & 46.47 & 37.3\\
	12 & 58.02 & 54.97\\
	13 & 87.46 & 86.53\\
	14 & 75.96 & 81.67\\
	15 & 53.86 & 69.2\\
	16 & 76.55 & 82.15\\
	17 & 52.67 & 67.1\\
	18 & 53.73 & 68.7\\
	19 & 105.36 & 114.16\\
	20 & 72.22 & 99.81\\
	21 & 88.97 & 59.78\\
	22 & 93.83 & 96.82\\
	23 & 97.13 & 79.15\\
	24 & 117.02 & 89.05\\
	25 & 99.13 & 107.06\\
	26 & 73.85 & 75.87\\
	27 & 70.5 & 85.45\\
	28 & 75.84 & 56.48\\
	29 & 102.93 & 71.41\\
	30 & 91 & 125.21\\
	31 & 110.55 & 87.39\\
	32 & 118.71 & 118.78\\
	33 & 96.14 & 129.86\\
	34 & 85.98 & 129.81\\
	35 & 122.12 & 86.05\\
	36 & 91.28 & 65.79\\
	37 & 109.94 & 149.36\\
	38 & 148.1 & 144.9\\
	39 & 171.1 & 167.08\\
	40 & 127.44 & 91.67\\
	41 & 137.29 & 99.69\\
	42 & 131.3 & 127.37\\
	43 & 152.85 & 188.63\\
	44 & 180.24 & 183.28\\
	45 & 135.05 & 164.04\\
	46 & 162.57 & 112.08\\
	47 & 179.86 & 184.16\\
	48 & 330.27 & 306.18\\
	49 & 493.7 & 413.12\\
	50 & 463.82 & 490.18\\
	51 & 410.82 & 438.25\\
	52 & 426.01 & 358.94\\
	53 & 383.01 & 426.58\\
	54 & 414.12 & 410.7\\
	55 & 386.58 & 405.27\\
	56 & 411.82 & 427.47\\
	57 & 474.55 & 365.41\\
	58 & 429.18 & 363.98\\
	59 & 298.68 & 260.8\\
	60 & 412.81 & 387.35\\
	61 & 494.82 & 492.73\\
	62 & 388.16 & 343.94\\
	63 & 367.72 & 320.76\\
	64 & 351.59 & 251.47\\
	65 & 308.13 & 339.59\\
	66 & 368.31 & 364.85\\
	67 & 402.94 & 329.38\\
	68 & 377.69 & 337.68\\
	69 & 377.71 & 332.63\\
	70 & 319.9 & 350.39\\
	71 & 343.88 & 351.11\\
	72 & 501.9 & 522.9\\
	73 & 341.41 & 288.43\\
	74 & 58.31 & 34.93\\
	75 & 9.13 & 14.15\\
	76 & 3.34 & 3.59\\
	77 & 170.4 & 754.48\\
	78 & 8.5 & 8.94\\
	1200 & 1200 & 1200\\
}\results

\section{Related Work}
\paragraph{Paracoherent Semantics.}
Many non-monotonic semantics for logic programs with negation have been proposed that can be considered as paracoherent semantics 
\cite{DBLP:journals/ngc/Przymusinski91,vang-etal-91,
DBLP:journals/amai/EiterLS97,DBLP:conf/lpkr/Seipel97,Balduccini03logicprograms,pere-pint-95,DBLP:journals/japll/AlcantaraDP05,DBLP:journals/logcom/GalindoRC08,osor-etal-08}.
However, \cite{DBLP:journals/ai/AmendolaEFLM16} have shown that only semi-stable semantics~\cite{inou-saka-95} and semi-equilibrium semantics~\cite{DBLP:journals/ai/AmendolaEFLM16} satisfy the following desiderata properties: 
(i) every consistent answer set of a program corresponds to a paracoherent answer set (\textit{answer set coverage});
(ii) if a program has some (consistent) answer set, then its paracoherent answer sets correspond to answer sets (\textit{congruence});
(iii) if a program has a classical model, then it has a paracoherent answer set (\textit{classical coherence});
(iv) a minimal set of atoms should be undefined (\textit{minimal undefinedness});
(v) every true atom must be derived from the program (\textit{justifiability}).

\paragraph{Computational aspects.}
Our approach to the computation of paracoherent answer sets is related to the computation of minimal models of propositional theories.
The first approaches were proposed for implementing circumscriptive reasoning (cfr.~\cite{DBLP:books/ox/LAI2Hb94}).
Later the attention shifted to the computation of minimal models of first-order clauses~\cite{DBLP:conf/tableaux/Niemela96,DBLP:conf/cade/HasegawaFK00}. 
\cite{DBLP:conf/tableaux/Niemela96} proposed a tableaux-based method where candidate models are generated and then tested for minimality.
\cite{DBLP:conf/cade/HasegawaFK00} proposed a method able to reduce minimality tests on candidate models. 
The usage of hyperresolution for minimal models of first-order clauses was presented in \cite{DBLP:journals/jar/BryY00} and implemented in Prolog. 
As observed in~\cite{Koshimura09} these approaches do not take profit of modern non-chronological-backtracking-based solving technology. 
This limit was overcome in \cite{Koshimura09} by an algorithm for computing minimal models of SAT formulas that is based on the same principle as the Minimize algorithm. 
The computation of minimal models of SAT formulae can be reduced to computing a Minimal Set over a Monotone Predicate (MSMP)~\cite{DBLP:journals/ai/JanotaM16}. Thus algorithms for MSMP such as those described in~\cite{DBLP:conf/cav/Marques-SilvaJB13} could be adapted (by properly taking into account the jump in computational complexity) for computing paracoherent answer sets.
Efficient polynomial algorithms for a subclass of positive CNF theories was proposed in~\cite{DBLP:journals/ai/AngiulliBFP14}.
That method cannot be applied directly to find a model of $\Pi$ that is minimal on the extension of $gap$ predicate.
The Split algorithm is similar to the algorithms employed for computing cautious consequences of ASP programs~\cite{DBLP:journals/tplp/AlvianoDR14} and backbones of SAT formulas~\cite{DBLP:journals/aicom/JanotaLM15}; nonetheless, to the best of our knowledge, it has no related counterpart in the literature concerning the computation of minimal models.


\section{Conclusion}

In this paper, we have tackled the problem of computing paracoherent
answer sets, namely semi-stable and stable-equilibrium models, which
has not been addressed so far. We have proposed a number of
algorithms relying on a program transformation with subsequent
calls to an answer set solver. We have conducted an experimental
analysis of these algorithms using incoherent programs of the ASP
competition, the analysis of incoherent answer set programs being the
prime application that we envision for paracoherent answer sets. The
experiments show that  algorithms Minimize and Split outperform other tested algorithms. The results also show
that the computation of a paracoherent answer set is a difficult
problem not just theoretically, but also in practice.

\section*{Acknowledgements}
This work was partially supported by the EU H2020 Marie Sk{\l}odowska-Curie grant agreement No 690974 ``MIREL'', MIUR within project ``SI-LAB BA2KNOW'', 
by MISE under project ``PIUCultura'', N. F/020016/01-02/X27, and by GNCS-INDAM.


\newpage

\bibliographystyle{abbrv}
\bibliography{aaai17}

\begin{thebibliography}{10}

\bibitem{DBLP:journals/japll/AlcantaraDP05}
J.~Alc{\^{a}}ntara, C.~V. Dam{\'{a}}sio, and L.~M. Pereira.
\newblock An encompassing framework for paraconsistent logic programs.
\newblock {\em J. Applied Logic}, 3(1):67--95, 2005.

\bibitem{DBLP:conf/lpnmr/AlvianoDLR15}
M.~Alviano, C.~Dodaro, N.~Leone, and F.~Ricca.
\newblock Advances in {WASP}.
\newblock In {\em {LPNMR} 2015}, pages 40--54, 2015.

\bibitem{DBLP:journals/tplp/AlvianoDR14}
M.~Alviano, C.~Dodaro, and F.~Ricca.
\newblock Anytime computation of cautious consequences in answer set
  programming.
\newblock {\em {TPLP}}, 14(4-5):755--770, 2014.

\bibitem{DBLP:journals/ai/AmendolaEFLM16}
G.~Amendola, T.~Eiter, M.~Fink, N.~Leone, and J.~Moura.
\newblock Semi-equilibrium models for paracoherent answer set programs.
\newblock {\em Artif. Intell.}, 234:219--271, 2016.

\bibitem{DBLP:conf/ijcai/AmendolaGLV16}
G.~Amendola, G.~Greco, N.~Leone, and P.~Veltri.
\newblock Modeling and reasoning about {NTU} games via answer set programming.
\newblock In {\em {IJCAI} 2016}, pages 38--45, 2016.

\bibitem{DBLP:journals/ai/AngiulliBFP14}
F.~Angiulli, R.~Ben{-}Eliyahu, F.~Fassetti, and L.~Palopoli.
\newblock On the tractability of minimal model computation for some {CNF}
  theories.
\newblock {\em Artif. Intell.}, 210:56--77, 2014.

\bibitem{Balduccini03logicprograms}
M.~Balduccini and M.~Gelfond.
\newblock Logic programs with consistency-restoring rules.
\newblock In {\em ISLFCR, AAAI 2003 Spring Symposium Series}, pages 9--18,
  2003.

\bibitem{DBLP:journals/cacm/BrewkaET11}
G.~Brewka, T.~Eiter, and M.~Truszczynski.
\newblock Answer set programming at a glance.
\newblock {\em Com. {ACM}}, 54(12):92--103, 2011.

\bibitem{DBLP:journals/jar/BryY00}
F.~Bry and A.~H. Yahya.
\newblock Positive unit hyperresolution tableaux and their application to
  minimal model generation.
\newblock {\em J. Autom. Reasoning}, 25(1):35--82, 2000.

\bibitem{DBLP:journals/tkde/BuccafurriLR00}
F.~Buccafurri, N.~Leone, and P.~Rullo.
\newblock Enhancing disjunctive datalog by constraints.
\newblock {\em {IEEE} Trans. Knowl. Data Eng.}, 12(5):845--860, 2000.

\bibitem{DBLP:journals/ai/CalimeriGMR16}
F.~Calimeri, M.~Gebser, M.~Maratea, and F.~Ricca.
\newblock Design and results of the fifth answer set programming competition.
\newblock {\em Artif. Intell.}, 231:151--181, 2016.

\bibitem{DBLP:journals/jetai/CampeottoDP15}
F.~Campeotto, A.~Dovier, and E.~Pontelli.
\newblock A declarative concurrent system for protein structure prediction on
  {GPU}.
\newblock {\em J. Exp. Theor. Artif. Intell.}, 27(5):503--541, 2015.

\bibitem{DBLP:journals/tplp/DodaroGLMRS16}
C.~Dodaro, P.~Gasteiger, N.~Leone, B.~Musitsch, F.~Ricca, and K.~Shchekotykhin.
\newblock {Combining Answer Set Programming and domain heuristics for solving
  hard industrial problems (Application Paper)}.
\newblock {\em {TPLP}}, 16(5-6):653--669, 2016.

\bibitem{DBLP:conf/rr/DodaroLNR15}
C.~Dodaro, N.~Leone, B.~Nardi, and F.~Ricca.
\newblock Allotment problem in travel industry: {A} solution based on {ASP}.
\newblock In B.~ten Cate and A.~Mileo, editors, {\em {RR} 2015}, volume 9209 of
  {\em LNCS}, pages 77--92. Springer, 2015.

\bibitem{DBLP:journals/amai/EiterLS97}
T.~Eiter, N.~Leone, and D.~Sacc{\`{a}}.
\newblock On the partial semantics for disjunctive deductive databases.
\newblock {\em Ann. Math. Artif. Intell.}, 19(1-2):59--96, 1997.

\bibitem{DBLP:books/ox/LAI2Hb94}
D.~M. Gabbay, C.~J. Hogger, J.~A. Robinson, and J.~H. Siekmann, editors.
\newblock {\em Handbook of Logic in Artificial Intelligence and Logic
  Programming, Volume2, Deduction Methodologies}.
\newblock Oxford University Press, 1994.

\bibitem{DBLP:journals/tplp/GagglMRWW15}
S.~Gaggl, N.~Manthey, A.~Ronca, J.~Wallner, and S.~Woltran.
\newblock Improved answer-set programming encodings for abstract argumentation.
\newblock {\em {TPLP}}, 15(4-5):434--448, 2015.

\bibitem{DBLP:journals/logcom/GalindoRC08}
M.~J.~O. Galindo, J.~R.~A. Ram{\'{\i}}rez, and J.~L. Carballido.
\newblock Logical weak completions of paraconsistent logics.
\newblock {\em J. Log. Comput.}, 18(6):913--940, 2008.

\bibitem{DBLP:series/synthesis/2012Gebser}
M.~Gebser, R.~Kaminski, B.~Kaufmann, and T.~Schaub.
\newblock {\em Answer Set Solving in Practice}.
\newblock Morgan {\&} Claypool Publishers, 2012.

\bibitem{DBLP:conf/lpnmr/GebserMR15}
M.~Gebser, M.~Maratea, and F.~Ricca.
\newblock The design of the sixth answer set programming competition - report
  -.
\newblock In {\em {LPNMR} 2015}, pages 531--544, 2015.

\bibitem{gelf-lifs-91}
M.~Gelfond and V.~Lifschitz.
\newblock Classical negation in logic programs and disjunctive databases.
\newblock {\em New Generation Comput.}, 9(3/4):365--386, 1991.

\bibitem{DBLP:conf/birthday/GrassoLMR11}
G.~Grasso, N.~Leone, M.~Manna, and F.~Ricca.
\newblock {ASP} at work: Spin-off and applications of the {DLV} system.
\newblock In {\em Logic Programming, Knowledge Representation, and Nonmonotonic
  Reasoning}, LNCS 6565, pages 432--451, 2011.

\bibitem{DBLP:conf/cade/HasegawaFK00}
R.~Hasegawa, H.~Fujita, and M.~Koshimura.
\newblock Efficient minimal model generation using branching lemmas.
\newblock In {\em CADE-17, 2000}, pages 184--199, 2000.

\bibitem{DBLP:journals/aicom/JanotaLM15}
M.~Janota, I.~Lynce, and J.~Marques{-}Silva.
\newblock Algorithms for computing backbones of propositional formulae.
\newblock {\em {AI} Commun.}, 28(2):161--177, 2015.

\bibitem{DBLP:journals/ai/JanotaM16}
M.~Janota and J.~Marques{-}Silva.
\newblock On the query complexity of selecting minimal sets for monotone
  predicates.
\newblock {\em Artif. Intell.}, 233:73--83, 2016.

\bibitem{Koshimura09}
M.~Koshimura, H.~Nabeshima, H.~Fujita, and R.~Hasegawa.
\newblock Minimal model generation with respect to an atom set.
\newblock In {\em {FTP} 2009}, {CEUR} 556, 2009.

\bibitem{DBLP:journals/tplp/MannaRT15}
M.~Manna, F.~Ricca, and G.~Terracina.
\newblock Taming primary key violations to query large inconsistent data via
  {ASP}.
\newblock {\em {TPLP}}, 15(4-5):696--710, 2015.

\bibitem{DBLP:conf/cav/Marques-SilvaJB13}
J.~Marques{-}Silva, M.~Janota, and A.~Belov.
\newblock Minimal sets over monotone predicates in boolean formulae.
\newblock In {\em {CAV} 2013}, pages 592--607, 2013.

\bibitem{DBLP:conf/tableaux/Niemela96}
I.~Niemel{\"{a}}.
\newblock A tableau calculus for minimal model reasoning.
\newblock In {\em {TABLEAUX} 1996}, pages 278--294, 1996.

\bibitem{osor-etal-08}
M.~Osorio, J.~R.~A. Ram\'{\i}rez, and J.~L. Carballido.
\newblock Logical weak completions of paraconsistent logics.
\newblock {\em J. Log. Comput.}, 18(6):913--940, 2008.

\bibitem{DBLP:journals/amai/Pearce06}
D.~Pearce.
\newblock Equilibrium logic.
\newblock {\em Ann. Math. Artif. Intell.}, 47(1-2):3--41, 2006.

\bibitem{pere-pint-95}
L.~M. Pereira and A.~M. Pinto.
\newblock Revised stable models - a semantics for logic programs.
\newblock In {\em EPIA}, pages 29--42, 2005.

\bibitem{DBLP:journals/ngc/Przymusinski91}
T.~C. Przymusinski.
\newblock Stable semantics for disjunctive programs.
\newblock {\em New Generation Comput.}, 9(3/4):401--424, 1991.

\bibitem{inou-saka-95}
C.~Sakama and K.~Inoue.
\newblock Paraconsistent stable semantics for extended disjunctive programs.
\newblock {\em J. Log. Comput.}, 5(3):265--285, 1995.

\bibitem{DBLP:conf/lpkr/Seipel97}
D.~Seipel.
\newblock Partial evidential stable models for disjunctive deductive databases.
\newblock In {\em LPKR}, pages 66--84, 1997.

\bibitem{vang-etal-91}
A.~{van Gelder}, K.~Ross, and J.~Schlipf.
\newblock The well-founded semantics for general logic programs.
\newblock {\em J.~ACM}, 38(3):620--650, 1991.

\end{thebibliography}

\label{lastpage}
\end{document}